\documentclass[useAMS,usenatbib]{mn2e}

\usepackage{graphicx}
\usepackage{natbib}
\usepackage[figuresleft]{rotating}

\newcommand{\ltsima} {$\; \buildrel < \over \sim \;$}
\newcommand{\gtsima} {$\; \buildrel > \over \sim \;$}
\newcommand{\lta} {\lower.5ex\hbox{\ltsima}}
\newcommand{\gta} {\lower.5ex\hbox{\gtsima}}
\newcommand{\kmsMp}{km s$^{-1}$\,Mpc$^{-1}$}
\newcommand{\kms}{km\ s$^{-1}$}

\newcommand{\lya}{Ly$\alpha$}
\title[CO(1-0) in the Spiderweb Galaxy]{CO(1-0) detection of molecular gas in the massive Spiderweb Galaxy ($z$=2)\thanks{From observations with the Australia Telescope Compact Array}}
\author[B. H. C. Emonts et al.]{B. H. C. Emonts$^{1}$\thanks{E-mail:bjorn.emonts@csiro.au}, I. Feain$^{1}$, H. J. A. R\"{o}ttgering$^{2}$, G. Miley$^{2}$, N. Seymour$^{1}$, R. P.
\newauthor Norris$^{1}$, C. L. Carilli$^{3}$, M. Villar-Mart\'{i}n$^{4}$, M. Y. Mao$^{3}$, E. M. Sadler$^{5}$, R.\,D. Ekers$^{1}$, 
\newauthor G.\,A. van Moorsel$^{3}$, R.\,J. Ivison$^{6,7}$, L. Pentericci$^{8}$, C.\,N. Tadhunter$^{9}$, D.\,J. Saikia$^{10,11}$\\
$^{1}$CSIRO Astronomy and Space Science, Australia Telescope National Facility, PO Box 76, Epping NSW, 1710, Australia\\
$^{2}$Leiden Observatory, University of Leiden, P.O. Box 9513, 2300 RA Leiden, Netherlands\\
$^{3}$National Radio Astronomy Observatory, P.O. Box 0, Socorro, NM 87801-0387, USA \\
$^{4}$Centro de Astrobiolog\'{i}a (INTA-CSIC), Ctra de Torrej\'{o}n a Ajalvir, km 4, 28850 Torrej\'{o}n de Ardoz, Madrid Spain\\
$^{5}$School of Physics, University of Sydney, NSW 2006, Australia\\
$^{6}$UK Astronomy Technology Centre, Science and Technology Facilities Council, Royal Observatory, Blackford Hill, Edinburgh EH9 3HJ\\
$^{7}$Institute for Astronomy, University of Edinburgh, Blackford Hill, Edinburgh EH9 3HJ\\ 
$^{8}$INAF Osservatorio Astronomico di Roma, Via Frascati 33,00040 Monteporzio (RM), Italy\\
$^{9}$Department of Physics and Astronomy, University of Sheffield, Sheffield S3 7RH, UK\\
$^{10}$Cotton College State University, Panbazar, Guwahati 781 001, India\\
$^{11}$National Centre for Radio Astrophysics, TIFR, Ganeshkhind, Pune 411 007, India
}
\begin{document}

\date{}

\pagerange{\pageref{firstpage}--\pageref{lastpage}} \pubyear{2010}

\maketitle

\label{firstpage}

\begin{abstract}
The high-redshift radio galaxy MRC\,1138-262 (`Spiderweb Galaxy'; $z = 2.16$), is one of the most massive systems in the early Universe and surrounded by a dense `web' of proto-cluster galaxies. Using the Australia Telescope Compact Array, we detected CO(1-0) emission from cold molecular gas -- the raw ingredient for star formation -- across the Spiderweb Galaxy. We infer a molecular gas mass of M$_{\rm H2} =  6 \times 10^{10}$ M$_{\odot}$ (for M$_{\rm H2}$/L'$_{\rm CO}$\,=\,0.8). While the bulk of the molecular gas coincides with the central radio galaxy, there are indications that a substantial fraction of this gas is associated with satellite galaxies or spread across the inter-galactic medium on scales of tens of kpc. In addition, we tentatively detect CO(1-0) in the star-forming proto-cluster galaxy HAE~229, 250 kpc to the west. Our observations are consistent with the fact that the Spiderweb Galaxy is building up its stellar mass through a massive burst of widespread star formation. At maximum star formation efficiency, the molecular gas will be able to sustain the current star formation rate (SFR $\approx$ 1400 M$_{\odot}$\,yr$^{-1}$, as traced by Seymour et al.) for about 40 Myr. This is similar to the estimated typical lifetime of a major starburst event in infra-red luminous merger systems.

\end{abstract}


\section{Introduction}
\label{sec:intro}

High-$z$ Radio Galaxies (HzRGs) are signposts of large over-densities in the early Universe, or proto-clusters, which are believed to be the ancestors of local rich clusters \citep[e.g.][]{mil08,ven07}. Historically, HzRGs were often identified by the ultra-steep spectrum of their easily detectable radio continuum, which served as a beacon for tracing the surrounding faint proto-cluster \citep{rot94,cha96}. 
HzRGs are typically the massive central objects in these proto-clusters. 

One of the most impressive HzRGs is MRC\,1138-262, also called the `Spiderweb Galaxy' \citep[$z=2.16$;][]{pen97,mil06}. It is one of the most massive galaxies in the early Universe \citep[M$_{\star}$$\sim$2$\times$10$^{12}$\,M$_{\odot}$;][]{sey07,bre10}. The Spiderweb Galaxy is a conglomerate of star forming clumps \citep[or `galaxies', following][]{hat09} that are embedded in a giant ($>$200\,kpc) \lya\ halo, located in the core of the Spiderweb proto-cluster \citep{pen97,car98,car02}. The central galaxy hosts the ultra-steep spectrum radio source MRC\,1138-262 \citep[$-1.2 \leq \alpha^{\rm 8.1\,GHz}_{\rm 4.5\,GHz} \leq -2.5$ for the various continuum components;][]{pen97,car97}. This source exerts dramatic feedback onto the \lya\ gas \citep{nes06}. 


Emission-line surveys identified tens of galaxies to be associated with the Spiderweb Galaxy and its surrounding proto-cluster \citep{pen00,kur04,cro05,doh10,kui11}. These galaxies harbour a significant fraction of the system's unobscured star formation \citep{hat09}.
Over-densities of red sequence galaxies were found by \citet{kod07} and \citet{zir08}. At Mpc scales, an overdensity of X-ray AGN trace a filamentary structure roughly aligned with the radio axis \citep{pen02}. \citet{kui11} showed that the proto-cluster is dynamically evolved and a possible merger of two subclusters. \citet{hat09} predicted that most of the central proto-cluster galaxies will merge with MRC\,1138-262 and double its stellar mass by $z=0$, but that gas depletion will have exhausted star formation long before, so that the Spiderweb Galaxy will evolve into a cD galaxy found in the centres of present-day clusters.

Sub-millimeter observations showed that the Spiderweb has massive star formation extended on scales of $>$200\,kpc \citep{ste03}, in agreement with PAH (Polycylic Aromatic Hydrocarbon) emission from MRC~1138-262 and two H$\alpha$-emitting companions \citep{ogl12}. Despite significant jet-induced feedback \citep{nes06,ogl12}, \citet{sey12} derived a high star formation rate of SFR $\approx$ 1400 M$_{\odot}$\,yr$^{-1}$ and an AGN accretion rate of 20$\%$ of the Eddington limit for MRC\,1138-262, indicating that it is in a phase of rapid growth of both black hole and host galaxy.

How long this phase will last and how much stellar mass will be added depends on the available fuel for the ongoing star formation. The raw ingredient for star formation (and potential AGN fuel) is molecular hydrogen (H$_{2}$). Extremely luminous mid-IR line emission from warm ($T>300$\,K), shocked H$_{2}$ gas has been detected in MRC\,1138-262 with {\sl Spitzer} \citep{ogl12}. This indicates that the radio jets may heat large amounts of molecular hydrogen, possibly quenching star formation in the nucleus (see \citealt{ogl12} for a discussion). However, in order to fuel the observed large star formation rate, an additional extensive reservoir of cold molecular gas must be present. An excellent tracer of the cold component of H$_{2}$ is carbon-monoxide, CO({\sl J,J-1}). Particularly efficient is the study of the ground-transition CO(1-0), which is the most robust tracer of the overall H$_{2}$ gas, including the widespread, low-density and sub-thermally excited component \citep[][]{pap00,pap01,pap02,dan09,car10,ivi11}.

In this paper,\,we present the detection of CO(1-0) in the Spiderweb Galaxy. We assume H$_{0} = 71$\,\kmsMp, $\Omega_{\rm M} = 0.27$ and  $\Omega_{\rm \Lambda} = 0.73$ (i.e. angular distance scale of 8.4 kpc arcsec$^{-1}$ and luminosity distance $D_{\rm L} = 17309$ Mpc).

\section{Observations}
\label{sec:observations}

CO(1-0) observations were performed with the Australia Telescope Compact Array (ATCA) during Aug 2011 - Mar 2012 in the compact hybrid H75 and H168 array configurations. The total on-source integration time was 22h \citep[after discarding data taken in poor weather, i.e. atmospheric path length rms fluctuations $>$\,400\,$\mu$m;][]{mid06}. Both 2\,GHz ATCA bands were centred close to $\nu_{\rm obs}$=36.5\,GHz (T$_{\rm sys}$$\sim$$70-100$K), corresponding to the redshifted CO(1-0) line. The phases and bandpass were calibrated every 5-15 min with a 2 min scan on the nearby bright calibrator PKS\,1124-186. Fluxes were calibrated using Mars.

For the data reduction we followed \citet{emo11b}. The relative flux calibration accuracy between runs was $\la$\,5\,$\%$, while the uncertainty in absolutely flux accuracy was up to $20\%$ based on the flux-model for Mars (version March 2012).
The broad 2\,GHz band ($\Delta$v $\approx 16,000$ \kms) allowed us to separate the continuum from the line emission in the uv-domain by fitting a straight line to the line-free channels. We Fourier transformed the line data\footnote{Similarly, we made a continuum map (10.39''$\times$6.55'', PA\,75.5$^{\circ}$). We detect the 36.6\,GHz radio continuum with an integrated flux of $S_{\rm 36.6\,GHz}$ = 10.7 mJy ($P_{\rm 36.6 GHz} = 3.6 \times 10^{26}$ W\,Hz$^{-1}$) across three beam-sizes, following the morphology of high resolution 4.7/8.2\,GHz data of \citet{car97}. A detailed discussion on the 36.6\,GHz radio continuum is deferred to a future paper.} to obtain a cube with robust weighting +1 \citep{bri95}, beam-size 9.54''\,$\times$\,5.31'' (PA\,63.3$^{\circ}$) and channel width 8.6 \kms. 
The line data were binned by 15 channels and subsequently Hanning smoothed to a velocity resolution of 259 \kms, resulting in a noise level of 0.085 mJy\,bm$^{-1}$ per channel. 

The spectra presented in this paper were extracted against the central pixel in the regions descibed in the text (pixelsize $2.3'' \times 2.3''$), unless otherwise indicated. Total intensity images of the CO(1-0) emission were made by summing the channels across which CO(1-0) was detected. All estimates of $L'_{\rm CO}$ in this paper have been derived from these total intensity images. The data were corrected for primary beam attenuation (FWHM$_{\rm PrimBeam} = 77$\,arcsec) and are presented in optical barycentric velocity with respect to $z=2.161$.

\section{Results}
\label{sec:results}

\begin{figure*}
\centering
\includegraphics[width=0.89\textwidth]{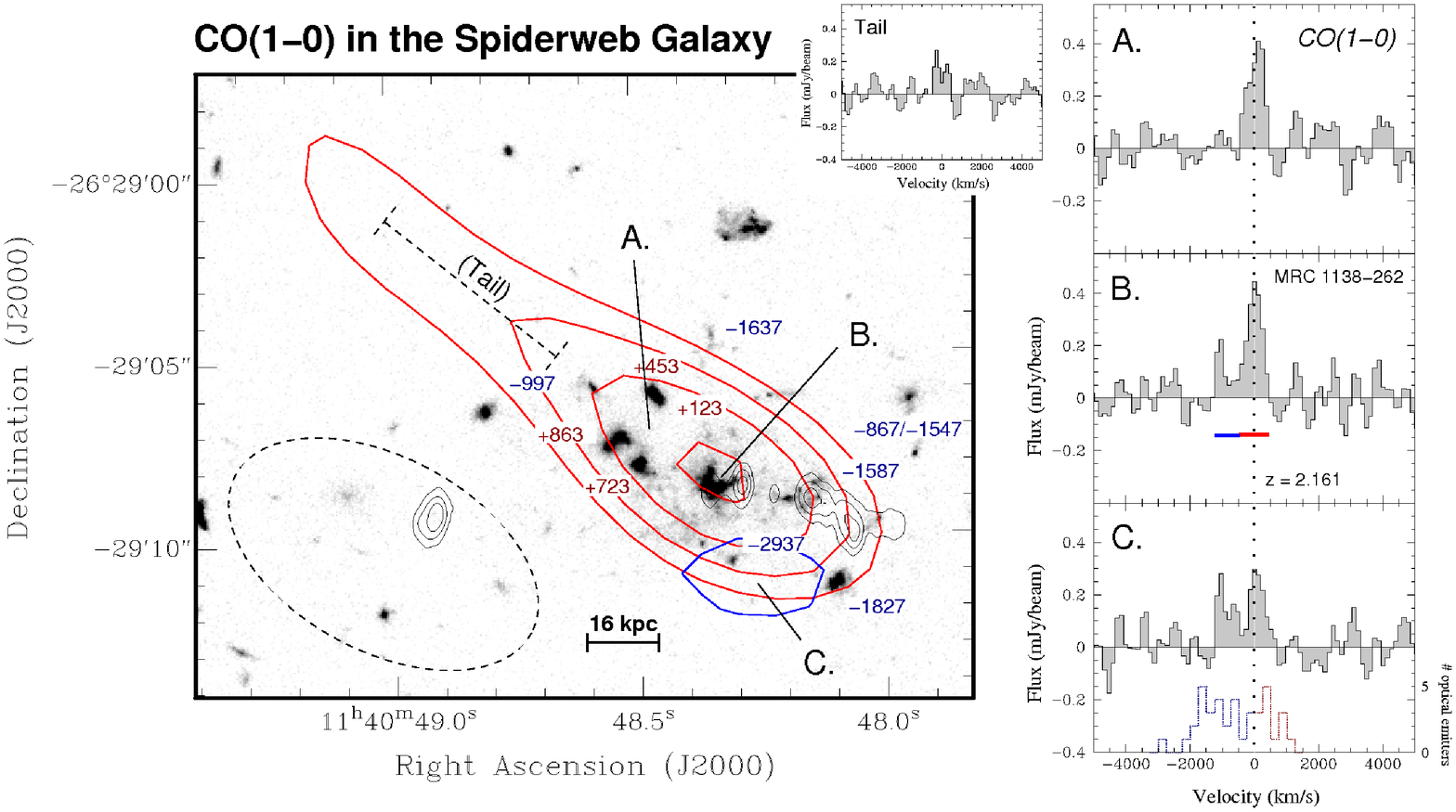}
\caption{{\sl Left:} Contours of the CO(1-0) emission overlaid onto an {\sl HST/ACS (g$_{475}$+I$_{814}$)} image of the Spiderweb Galaxy \citep[][]{mil06}. The plot roughly equals the size of the giant \lya\ halo (see Fig.\,\ref{fig:HAE229}), hence the individual galaxies shown here are part of the Spiderweb Galaxy. Contour levels of the CO(1-0) emission are: 2.8, 3.5, 4.2, 5.0\,$\sigma$ (no negative contours are present at the same -2.8$\sigma$ level, but see Fig.\,\ref{fig:HAE229} for a larger field-of-view). The red and blue contours represent the red and blue part of the double-peaked CO(1-0) profile, as indicated by the horizontal bar in the right-middle plot (velocity ranges: red: $ -453 < {\rm v} < +453$\,\kms\ with $\sigma = 0.046$ Jy\,bm$^{-1}$$\times$\kms; blue: $-1228 < {\rm v} < -453$\,\kms\ with $\sigma = 0.043$ Jy\,bm$^{-1}$$\times$\kms). Note that the blue signal is too weak to be detected across the FWHM of the sythesized beam, hence the emission and its exact location need to be verified with future observations. The black contours represent the 8.2\,GHz radio continuum from \citet{car97}. The dashed ellipse in the bottom-left corner shows the ATCA beam-size. The red and blue numbers indicate the velocities of individual galaxies as derived from optical emission lines \citep[adjusted for $z=2.161$;][]{kui11}. {\sl Right:} CO(1-0) emission-line profiles at four different regions within the Spiderweb Galaxy. The apertures of regions A, B and C are described in Sect \ref{sec:observations}. The spectrum of the tentative NE tail was taken by averaging 4 pixels along the dashed line in the left plot. Although the profiles are not mutually independent, they suggest that there is a change in the CO(1-0) kinematics across the regions A, B and C. The velocity is centred on $z=2.161$ (Sect. \ref{sec:results}). The dotted histogram in the bottom plot shows the distribution of optical and UV rest-frame line emitters in the Spiderweb Galaxy and surrounding proto-cluster \citep[from][]{kui11}. {\sl [See the electronic version of the paper for color images.]}}
\label{fig:map}
\end{figure*}

We detect CO(1-0) emission in the Spiderweb Galaxy (Fig. \ref{fig:map}). The CO(1-0) profile appears double-peaked, with a firm 5$\sigma$ `red' peak and tentative 3$\sigma$ `blue' peak, separated by $\sim$1000 \kms (with $\sigma$ derived from the integrated line profile). Figure \ref{fig:map} {\sl (left)} shows a total intensity map of the red and blue component. The total (`red+blue') CO(1-0) emission-line luminosity that we derive is $L'_{\rm CO} = 7.2 \pm 0.6 \times 10^{10}$\, ${\rm K~km~s^{-1}~pc^2}$ \citep[following equation 3 in][]{sol05}.\footnote{The measurement error in $L'_{\rm CO}$ does not include a 20$\%$ uncertainty in the model of our used flux calibrator Mars (Sect. \ref{sec:observations}).} Table \ref{tab:table} summarises the CO(1-0) emission line properties.

%

Figure\,\ref{fig:map} {\sl (left)} shows that the bulk of the CO(1-0) coincides with the radio galaxy (region `B'). However, there are strong indications from both the gas kinematics and distribution that a significant fraction of the CO(1-0) emission is spread across tens of kpc. 

\begin{figure*}
\centering
\includegraphics[width=\textwidth]{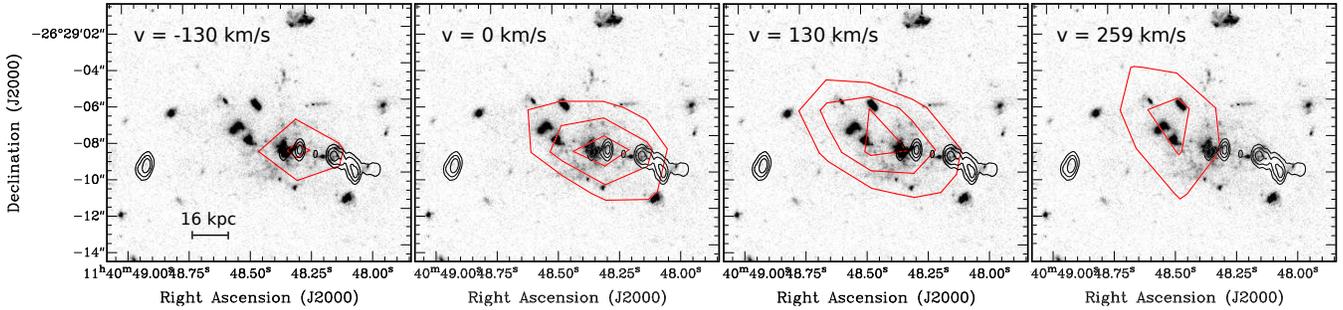}
\caption{Channels maps of the CO(1-0) emission in regions A and B, highlighting a gradient in the gas kinematics. Contour levels of the CO(1-0) emission (red) are: 3.0, 3.8, 4.5$\sigma$ (with $\sigma = 0.085$\,mJy\,bm$^{-1}$; Sect. \ref{sec:observations}). No negative features are present in these channels at the same level (-3$\sigma$). The radio continuum (black contours), beam-size and redshift are the same as in Fig.\,\ref{fig:map}.}
\label{fig:chanmaps}
\end{figure*}

First, despite the limited spatial resolution of our observations, there appears to be a velocity gradient in the gas kinematics across the inner 30-40 kpc of the Spiderweb. As shown in Fig.\,\ref{fig:map}, the redshift of the CO(1-0) peak emission decreases when going from region `A' to region `B' to region `C'. Most prominent is the apparent spatial separation between the peak of the blue component of the double-peaked CO(1-0) profile in region C and the peak of the red component in region B. Fig.\,\ref{fig:chanmaps} visualises that also between regions B and A there is a clear velocity gradient in the CO(1-0) emission. The decrease in the velocity of the CO(1-0) peak emission from region A $\rightarrow$ B $\rightarrow$ C is consistent with a decrease in redshift of optical line emitting galaxies found in these regions (see Fig.\,\ref{fig:map}, though note that the large ATCA beam prevents us from spatially resolving the individual galaxies in our CO data). Fig.\,\ref{fig:map} also shows indications for an extension (`tail') of the CO(1-0) emission beyond region A (stretching up to 100 kpc NE of the radio core), but this tail is detected only at a 3$\sigma$ level and thus needs to be verified. We have started an observational program to verify the distribution and kinematics of the CO(1-0) emission at higher sensitivity and spatial resolution (results will be reported in a future paper). 

In addition, the double-peaked CO(1-0) profile spreads over 1700 \kms\ (FWZI). This is extreme compared to what is found for quasars and submm-galaxies \citep[see][and references therein]{cop08,wan10,ivi11,rie11,bot12,kri12}. A few notable exceptions are high-$z$ systems in which the broad CO profiles arise from merging galaxies \citep[][and references therein]{sal12}. As can be seen in Fig.\,1 {\sl (bottom right)}, the double-peaked CO profile resembles the velocity distribution of optical line emitters detected in the Spiderweb proto-cluster \citep{kui11}, be it with a lower velocity dispersion of the CO gas, in particular on the blue-shifted side. 

These results thus indicate that a significant fraction of the CO(1-0) detected in the Spiderweb Galaxy likely originates from (merging) satellites of the central radio galaxy, or the inter-galactic medium (IGM) between them.

\begin{table*}
\caption{CO(1-0) properties in the various regions of the Spiderweb Galaxy and HAE\,229: $z_{\rm CO(1-0)}$=redshift, v$_{\rm CO(1-0)}$=velocity of the peak emission w.r.t $z_{\rm CO(1-0)}$=2.161; FWHM=CO(1-0) line width at half the maximum intensity; FWZI=full width of the CO(1-0) profile at zero intensity; $S_{\nu}$\,(peak)=CO peak flux; $I_{\rm CO(1-0)}$=CO integrated flux; $L'_{\rm CO(1-0)}$=CO total intensity: M$_{\rm H2}$=H$_{\rm 2}$ mass (Sect. \ref{sec:mass}). $z$, v, FWHM and $S_{\nu}$ have been derived from Gaussian fits to the CO profiles in Figs. \ref{fig:map} and \ref{fig:HAE229}. $I_{\rm CO(1-0)}$, $L'_{\rm CO(1-0)}$ and M$_{\rm H2}$ have been derived from the total intensity images in Figs. \ref{fig:map} and \ref{fig:HAE229}.}
\label{tab:table}
\vspace{-0.3mm}
\begin{tabular}{llcc|lll|ccl}
\hline
 & & & & \multicolumn{3}{c}{\bf Spiderweb Galaxy (SG)} & & & {\bf HAE\,229} \\
 & & & & SG region A & SG region B & SG region C & & & \\
\hline
$z_{\rm CO(1-0)}$ & & & & 2.163 $\pm$ 0.001 & 2.161 $\pm$ 0.001 & 2.150 $\pm$ 0.001 & & & 2.147 $\pm$ 0.001 \\
v$_{\rm CO(1-0)}$ & (\kms) &  & & 175 $\pm$ 75 & 0 $\pm$ 30 & -1060$^{+185}_{-40}$ & & & -1355 $\pm$ 65 \\
FWHM & (\kms) & & & 550$^{+165}_{-210}$$^{\dagger}$ & 540 $\pm$ 65 & 550$^{+150}_{-300}$$^{\dagger}$ & & & 395 $\pm$ 75 \\
FWZI & (\kms) & &  & \multicolumn{2}{l}{\hspace{6mm}905 $\pm$ 130 (A\,+\,B)$^{\ddagger}$} & 775 $\pm$ 130 & & & 520 $\pm$ 130 \\
$S_{\nu}$(peak) & (mJy\,beam) & & & 0.44 $\pm$ 0.06 & 0.44 $\pm$ 0.06 & 0.30 $\pm$ 0.09 & & & 0.32 $\pm$ 0.05 \\
$I_{\rm CO(1-0)}$ & (Jy\,\kms) & & & \multicolumn{2}{l}{\hspace{6mm}0.28 $\pm$ 0.03 (A\,+\,B)$^{\ddagger}$} & 0.03 $\pm$ 0.01 & & & 0.14 $\pm$ 0.01 \\
${L'}_{\rm CO(1-0)}$ & ($\times$ 10$^{10}$ K\,\kms\,pc$^{2}$) & &  & \multicolumn{2}{l}{\hspace{6mm}6.5 $\pm$ 0.6 (A\,+\,B)$^{\ddagger}$} & 0.7 $\pm$ 0.2 & & & 3.3 $\pm$ 0.2 \\
M$_{\rm H2}$ & ($\times$ 10$^{10}$ M$_{\odot}$) & & & \multicolumn{2}{l}{\hspace{6mm}5 $\pm$ 1 (A\,+\,B)$^{\ddagger}$} & 0.6 $\pm$ 0.2 & & & 3 $\pm$ 1 \\
\hline
\end{tabular} 
\flushleft
$^{\dagger}$ Values are quoted for a single Gaussian profile fit, with errors reflecting uncertainties due to assymmetry of the corresponding profile component.\\
$^{\ddagger}$ Regions A and B are spatially unresolved and only marginally resolved kinematically, hence a single value is derived from the entire `red' part of the total intensity image of Fig.\,\ref{fig:map}.
\end{table*}

Our results also suggests that the redshift of the central radio galaxy is associated with the red peak of the CO(1-0) profile, giving $z_{\rm CO(1-0)} = 2.161 \pm 0.001$. \citet{kui11} discuss that determining the redshift from optical and UV rest-frame emission lines is bound to a much larger uncertainty, but they derive $2.158 < z < 2.170$, which is in agreement with our estimated $z_{\rm CO(1-0)}$.

\subsection{HAE~229}

Fig.\,\ref{fig:HAE229} shows that also the dusty star-forming galaxy HAE 229 \citep[M$_{\star}$\,$\sim$\,5\,$\times$\,10$^{11}$\,M$_{\odot}$;][]{kur04,doh10} is detected in CO(1-0) at 3.7$\sigma$ significance. We derive $L'_{\rm CO} = 3.3 \pm 0.2 \times 10^{10}\ {\rm K~km~s^{-1}~pc^2}$ for HAE\,229. Table \ref{tab:table} summarises the CO(1-0) properties. HAE\,229 is located 250 kpc (30'') west of MRC\,1138-262, i.e. outside the gaint \lya\ halo. The CO(1-0) signal peaks at $v = -1354$\,\kms, which agrees with the H$\alpha$ redshift from \citet[][]{kur04}. None of the other line-emitting galaxies outside the \lya\ halo, but within the field-of-view of our observations, is reliably detected in CO(1-0).

\section{Discussion}
\label{sec:discussion}

\subsection{Molecular gas in the Spiderweb}
\label{sec:mass}

We can estimate the mass of molecular gas by adopting a standard conversion factor $\alpha_{\rm x} = {\rm M}_{\rm H2}/{\rm L}'_{\rm CO} = 0.8$ [M$_{\odot}$ (K \kms\ pc$^{2}$)$^{-1}$] \citep[where M$_{\rm H2}$ includes a helium fraction; e.g.][]{sol05}. This is consistent with $\alpha_{\rm x}$ found in ultra-luminous infra-red galaxies \citep[$L_{\rm IR} > 10^{12} L_{\odot}$;][]{dow98}, but we stress that the conversion from $L'_{\rm CO}$ into M$_{\rm H2}$ is not yet well understood \citep[][]{tac08,ivi11} and that $\alpha_{\rm x}$ crucially depends on the properties of the gas, such as metallicity and radiation field \citep[][]{glo11}.
Adopting $\alpha_{\rm x} = 0.8$ M$_{\odot}$ (K \kms\ pc$^{2}$)$^{-1}$ results in an estimated molecular gas mass in the Spiderweb Galaxy of M$_{\rm H2} \sim 6 \times 10^{10}$ M$_{\odot}$. We argue that this is likely a conservative estimate, based on the adopted conversion factor and the fact that a large amount of shock-heated molecular gas resides in the warm (T\,$>$\,100\,K) phase \citep[see][]{ogl12}. The putative H$_{2}$ mass of HAE\,229 is M$_{\rm H2}$\,$\sim$\,3$\times$$10^{10}$\,M$_{\odot}$.

\subsubsection{Nature of the molecular gas}
\label{sec:nature}

In Sect. \ref{sec:results} we saw that, while the CO(1-0) distribution is concentrated on the central radio galaxy, the CO emission spreads across the inner 30-40 kpc (a region which is rich in satellite galaxies). Based on its distribution and kinematics, we argued that part of the molecular gas is thus most likely associated with these satellite galaxies or the IGM between them. The extreme FWZI of the CO(1-0) emission (Sect. \ref{sec:results}) is another indication that the double-peaked profile is not likely caused by a nuclear disc in the central radio galaxy. This is consistent with earlier speculation by \citet{ogl12} that the high star formation rates could be the result of the accretion of gas or gas-rich satellites (which were found by \citealt{hat09} to contain most of the dust-uncorrected, instantaneous star formation), while nuclear star formation may be quenched by jet-induced heating of the molecular gas.

\begin{figure*}
\centering
\includegraphics[width=\textwidth]{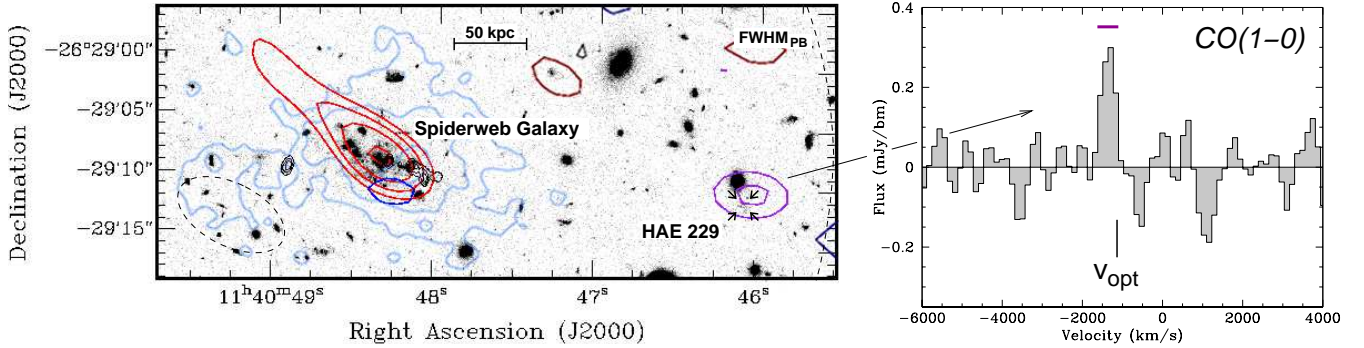}
\caption{
{\sl Left:} The Spiderweb Galaxy and HAE\,229. Shown is a zoom-out of Fig.\,\ref{fig:map} {\sl (left)} including the CO(1-0) contours of HAE\,229 (purple) 250\,kpc to the west. The light-blue contours indicate the extent of the \lya\ halo that encompasses the Spiderweb Galaxy (from \citealt{mil06}, see also \citealt{pen97}). The red and blue contours are the same as in Fig.\,\ref{fig:map} (2.8, 3.5, 4.2, 5.0$\sigma$), with $\sigma$ the noise level at the phase center (centred on the Spiderweb Galaxy). The dark-red and dark-blue contours are the corresponding negative signal (present only at -2.8$\sigma$ at significant distance from the phase center, where the noise increases due to correction of the signal for primary beam attennuation). Contour levels of the CO(1-0) in HAE\,229 are: -2.8 (grey), 2.8, 3.5$\sigma$ (purple), derived across te velocity range indicated by the purple bar in the spectrum on the right ($-1618 < {\rm v} < -1100$\,\kms; $\sigma = 0.048$ Jy\,bm$^{-1}$\,$\times$\,\kms). For HAE\,229, $\sigma$ is the local rms noise (i.e. 30 arcsec from the phase center) in the primary-beam corrected total intensity CO image. The dashed partial circle on the right shows the FWHM of the primary beam (i.e. the effective field-of-view) of our observations. The small arrows indicate the location of a peak in 24$\mu$m emission that coincides with HAE\,229 (data from \citealt{bre10}). {\sl Right:} CO(1-0) spectrum of HAE~229. The systemic velocity, as derived from optical emission lines, is also indicated \citep{kur04}. Note that the higher noise in this spectrum compared to the spectra of Fig.\,\ref{fig:map} is a result of the primary beam correction.}
\label{fig:HAE229}
\end{figure*}

The tentative NE tail (see Fig.\,\ref{fig:map}) spreads further out, beyond region A (which is rich in satellite galaxies) into a region with no known companion galaxies or detectable \lya\ emission (Fig. \ref{fig:HAE229}). However, a Mpc-scale filamentary structure exists in east-west direction \citep{pen02}. We speculate that -- if confirmed -- this tentative tail might indicate that cold gas is found, or being accreted along, this filament.

Two alternative scenarios that should be considered to explain the CO characteristics are AGN driven outflows and cooling flows. Cooling flows have been detected in CO in giant central cluster (cD) galaxies at $z$\,$<$\,0.4, some of which contain H$_{2}$ masses similar to that of the Spiderweb Galaxy. However, compared to the Spiderweb Galaxy, these cooling flow galaxies show much narrower typical line widths (FWHM$_{\rm CO} < 500$\,\kms, taking into account an uncertain $\alpha_{\rm x}$ and the use of narrow-band receivers; \citealt{edg01,sal03,sal06}). Radio-jet driven outflows of optical emission-line gas were found on scales of tens of kpc in the Spiderweb Galaxy by \citet{nes06}. Similar to the CO distribution, these optical emission lines are significantly more redshifted NW compared to SE of the radio core, though the ambiguity in optical redshifts makes a direct comparison difficult. Still, there is an interesting alignment between the redshifted CO(1-0) emission in region A and the region in which \citet{nes06} detect the fastest redshifted outflow velocities in the optical emission-line gas (their `zone 1', which stretches in between region A and B). The FWHM of the optical emission-line gas is, however, significantly larger than that of the CO(1-0) emission, indicating that it has a much larger velocity dispersion. Both the cooling flow and the radio-jet feedback scenario deserve further investigation, once CO observations with higher resolution and sensitvity have confirmed the extent of the CO(1-0) emission.

\subsection{Evolutionary stage}
\label{sec:evolution}

From fitting the mid- to far-IR spectral energy distribution, \citet{sey12} derive a starburst IR-luminosity of L$_{\rm IR} = 8 \times 10^{12}$\,L$_{\odot}$ and star formation rate of SFR\,=\,1390 M$_{\odot}$\,yr$^{-1}$ for MRC\,1138-262. L$_{\rm IR}$/L'$_{\rm CO(1-0)}$ agrees well with correlations found in various types of low- and high-$z$ objects \citep[e.g.][]{ivi11}. Assuming that all the H$_{2}$ is available to sustain the high SFR, we derive a minimum mass depletion time-scale of t$_{\rm depl} = \frac{{\rm M}_{\rm H2}}{\rm SRF} \approx 40$ Myr. This is comparable to the estimated typical lifetime of a major starburst episode in IR-luminous merger systems \citep{mih94,swi06}, though the bulk of the intense star formation in the Spiderweb Galaxy may occur on scales of tens of kpc (Sect. \ref{sec:nature}). The mass depletion time-scale may be shorter if the cold molecular gas is more rapidly depleted by feedback processes, such as shock-heating \citep{ogl12} or jet-induced outflows (found to occur at rates of $\sim$400\,M$_{\odot}$\,yr$^{-1}$ in the optical emission line gas by \citealt{nes06}). Nevertheless, both the current large star formation rate and cold molecular gas content imply that we are witnessing a phase of rapid galaxy growth though massive star formation, coinciding with the AGN activity.

The H$_{2}$ mass is much larger than the estimated mass of the emission-line gas in the \lya\ halo \citep[M$_{\rm emis} = 2.5 \times 10^{8}$\,M$_{\odot}$;][]{pen97}. However, \citet{car02} show that the radio source is enveloped by a region of hot, shocked X-ray gas of potentially M$_{\rm hot} = 2.5 \times 10^{12}$\,M$_{\odot}$. This suggests that, even when the current reservoir of cold molecular gas is consumed, there is a potential gas reservoir available for future episodes of starburst (and AGN) activity, provided that the gas can cool down to form molecular clouds \citep[e.g.][]{fab94} and this process is not entirely counter-acted by ongoing AGN feedback \citep{car02,nes06,ogl12}. The merger of proto-cluster galaxies with the Spiderweb Galaxy may also trigger a new burst of star formation, depending on the available gas reservoir in these systems. 

For the dusty star-forming galaxy HAE\,229, $L'_{\rm CO(1-0)}$ is comparable to that of IR-selected massive star-forming galaxies at $z$=1.5 \citep{ara10} and some high-$z$ submm galaxies \citep[e.g.][]{ivi11}. Our CO results are consistent with observations by \citet{ogl12} that HAE\,229 is going through a major and heavily obscured starburst episode. From their calculated SFR$\sim$880\,M$_{\odot}$\,yr$^{-1}$, we derive t$_{\rm depl} \approx 30$ Myr, i.e. similar to that of the Spiderweb Galaxy. 

\subsection{CO(1-0) in HzRGs}
\label{sec:coinhzrg}

MRC\,1138-262 is part of an ATCA survey for CO(1-0) in a southern sample of HzRGs \citep[$1.4<z<3$; Emonts et al in prep, see also][]{emo11a,emo11b}. So far, it is one of only very few secure CO(1-0) detections among HzRGs; two other examples being MRC\,0152-209 \citep[$z=1.92$;][]{emo11a} and 6C\,1909+72 \citep[$z=3.53$;][]{ivi12}. 

CO detections in HzRGs made with narrow-band receivers and/or higher transitions are also still limited in number \citep[][also review by \citealt{mil08}]{sco97,all00,pap00,pap01,bre03,bre03AR,bre05,gre04,kla05,nes09,emo11b,ivi08,ivi12}. However, in some cases CO is resolved on tens of kpc scales \citep{ivi12}, associated with various components \citep[e.g. merging gas-rich galaxies;][]{bre05}, or found in giant Ly$\alpha$ halos that surround the host galaxy \citep{nes09}. This shows that detectable amounts of cold molecular gas in HzRGs are not restricted to the central region of the radio galaxy. \citet{ivi12} discuss that CO detected HzRGs (often pre-selected on bright far-IR emission from a starburst) are generally associated with merger activity. This is also the case for the Spiderweb Galaxy, which is located in an extreme merger environment.

Using the JVLA, \citet{car11} mapped CO(2-1) emission throughout a $z$\,=\,4 proto-cluster associated with the sub-millimeter galaxy GN20 (tracing a combined mass of M$_{\rm H2} \sim 2 \times 10^{11}$\,M$_{\odot}$). Our results on the Spiderweb Galaxy and HAE\,229 are another example of the potential for studying the lowest CO transitions in proto-cluster environments with the ATCA and JVLA.

\section{Conclusions}
\label{sec:conclusions}

 We detect CO(1-0) emission from cold molecular gas across the massive Spiderweb Galaxy, a conglomerate of star forming galaxies at z=2.16. While the bulk of the CO(1-0) coincides with the central radio galaxy, part of the molecular gas is spread across tens of kpc. We explain that this gas is most likely associated with satellites of the central radio galaxy, or the IGM between them (though other scenarios are briefly discussed). The extensive reservoir of cold molecular gas likely provides the fuel for the widespread star formation that has been observed across the Spiderweb Galaxy. Continuous galaxy-merger and gas-accretion processes are the likely triggers for the observed high star formation rates. The total mass of cold gas ($M_{\rm H2} =  6 \times 10^{10}$$_{\,[\alpha_x=0.8]}$ M$_{\odot}$) is enough to sustain the current high star formation rate in the Spiderweb Galaxy for $\sim$\,40 Myr, which is similar to the typical lifetime of major starburst events seen in IR-luminous merger systems. Our CO results on the Spiderweb Galaxy show the potential for studying the cold gas throughout high-z proto-clusters with the ATCA, JVLA and ALMA.

\section*{Acknowledgments}
We thank Ernst Kuiper for sharing his HST imaging and the anonymous referee for useful feedback that improved this paper. We thank the Narrabri observatory staff for their help. BE acknowledges the Centro de Astrobiolog\'{i}a/INTA for their hospitality. NS is recipient of an ARC Future Fellowship. The Australia Telescope is funded by the Commonwealth of Australia for operation as a National Facility managed by CSIRO.


\end{document}